\documentstyle[aps,prb,multicol,epsfig]{revtex}
\begin{document}
\title{
Magnon dispersions in quantum Heisenberg ferrimagnetic chains\\
at zero temperature}
\author{N. B. Ivanov\cite{paddress} }
\address{
Theoretische Physik II, Universit\"at Augsburg,\\
D-86135 Augsburg, Germany
}
\date{\today}
\maketitle
\begin{abstract}
Within the Dyson-Maleev boson formalism,
we study the zero-temperature magnon dispersions
in a family of one-dimensional quantum
Heisenberg ferrimagnets
composed of two different spins $(S_1,S_2)$ in the
elementary cell. It is shown that
the spin-wave theory can produce
precise quantitative results for
the low-energy excitations.
The spin-stiffness constant $\rho_s$ and the
optical magnon gap $\Delta$ of different $(S_1,S_2)$
ferrimagnetic systems are calculated, respectively,  to second and
third order  in the quasiparticle
interaction. The  spin-wave
results are compared with available numerical estimates.
\end{abstract}
\draft
\begin{multicols}{2}
\section{Introduction}
\begin{figure}
\centering\epsfig{file=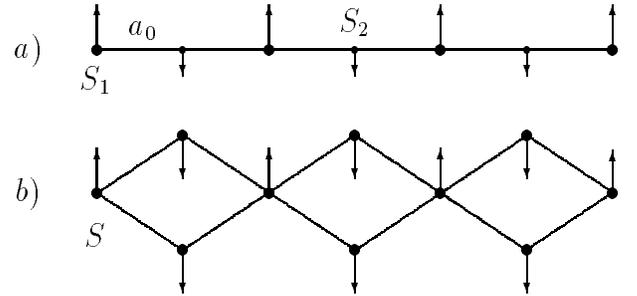,width=8.6cm}
\vspace{0.3cm}
\narrowtext
\caption{
Two types of
ferrimagnetic chains on bipartite lattices: a) mixed-spin
chain with alternating site spins $S_1$ and $S_2$ ($S_1>S_2$);
b) uniform-spin diamond chain with a site spin $S$.
}
\label{chains}
\end{figure}

In the last decade a large variety of
quasi-one-dimensional
(1D) mixed-spin  compounds with ferrimagnetic properties
has been  synthesized.\cite{pei} Most of them
are  molecular
magnets containing two different transition-metal magnetic
ions which are alternatively distributed
on the lattice (Fig.~\ref{chains}).
For example, two families of such compounds read
ACu(pba)(H$_2$O)$_3\cdot$nH$_2$O
and ACu(pbaOH)(H$_2$O)$_3\cdot$nH$_2$O,
where pba = 1,3-propylenebis (oxamato),
pbaOH = 2-hydroxo-1,3-propylenebis (oxamato),
and A=Ni, Fe, Co, and Mn.
Published experimental work
implies that the magnetic properties of these mixed-spin materials
are basically  described by a quantum Heisenberg spin model
with antiferromagnetically coupled
nearest-neighbor localized spins:
\begin{eqnarray}\label{ham}
{\cal H}&=& J\sum_{n =1}^{N}\left[
{\bf S}_1(n)+{\bf S}_1(n+1)\right]\cdot {\bf S}_2(n)\nonumber \\
&-&\mu_BH\sum_{n =1}^{N}\left[ g_1S_1^{z}(n)
+g_2S_2^{z}(n)\right],\hspace{0.1cm} J>0\, .
\end{eqnarray}
Here the integers $n$ number the $N$ elementary cells, each of them
containing two site spins with quantum
numbers $S_1>S_2$. The $g$-factors, related to the spins $S_1$ and
$S_2$,
are denoted by $g_1$ and $g_2$, respectively.
$\mu_B$ is the Bohr magneton,  and $H$
is the external uniform magnetic field
applied along the $z$ direction.
As an example, the following  values for the parameters in
Eq.\ (\ref{ham}) have been extracted
from magnetic measurements on the
recently synthesized quasi-1D  bimetallic  compound
NiCu(pba)(D$_2$O)$_3\cdot$2D$_2$O:
$(S_1,S_2)\equiv (S_{Ni},S_{Cu})=(1,\frac{1}{2})$,
$J/k_B=121K$, $g_1 \equiv g_{Ni}=2.22$, $g_2
\equiv g_{Cu}=2.09$.\cite{hagiwara}
In view of the  recent developments in the physics of uniform
quasi-1D systems (concerning, in particular, the
behavior of antiferromagnetic chains and ladders in external magnetic
fields: see, e.g., Ref.\ \onlinecite{giamarchi} and references therein),
ferrimagnetic chains and ladders open an interesting  new area.
The point is that the presence of two or more
different quantum  spins
in the elementary cell considerably
increases the number of situations of interest: the topology of
spin arrangements plays an essential role in the structure of
the ground state and low-energy excitations.\cite{fukui1}
For example, the diamond lattice  in Fig.~\ref{chains}
represents another interesting
class of Heisenberg ferrimagnetic chains constructed of one kind of
antiferromagnetically coupled site spins
but with a different number of lattice sites in the $\cal A$ and
$\cal B$ sublattices.\cite{macedo}
It is remarkable that only in the last few years
interest in the physical
properties of  1D quantum
Heisenberg ferrimagnets has considerably
increased:\cite{ivanov1} most of the early efforts
were concentrated on the
chemistry of molecular magnets and a relatively small amount
of work was devoted to physical properties.\cite{drillon1}

Since  interactions in the models presented in Fig.~\ref{chains}
set up  bipartite lattices, the Lieb-Mattis
theorem\cite{lieb} is applicable. In particular,
for the mixed-spin system it predicts
the existence of a ferrimagnetic ground state with
the total-spin quantum number
$S_g=(S_1-S_2)N$.
Thus the  model (\ref{ham}) exhibits
long-range ordered magnetic ground
state characterized by the magnetization density
$M_0=(S_1-S_2)/2a_0$,
$a_0$ being the lattice constant.
Such a magnetic phase may be referred to as {\em
a quantized unsaturated ferromagnetic phase}:
it is characterized by both
the quantized ferromagnetic order parameter $\bf M$
(quantized in integral or half-integral
multiples of the number of elementary cells, $M=M_0N$) and
the macroscopic sublattice  magnetizations
${\bf M}_A=\sum_{n=1}^N\langle {\bf S}_1(n)\rangle$
and ${\bf M}_B=\sum_{n=1}^N\langle {\bf S}_2(n)\rangle$.
What makes such a state and the
related transitions to magnetically disordered states
interesting is the fact that the ferromagnetic order parameter
is a conserved quantity:  in Heisenberg systems
this is expected to imply
strong constraints on the critical behavior.\cite{sachdev1}
Consequences of the spontaneous symmetry breaking
$SO(3)\rightarrow SO(2)$ (related to the establishment
of a ferrimagnetic ground state)
for the structure of low-energy excitations are dictated by
the nonrelativistic version of Goldstone's theorem\cite{nielsen}:
in the absence of long-range forces, a spontaneous
symmetry breaking leads to low-energy excitations
whose energy tends to zero
for wave-vectors  $k \rightarrow 0$. In contrast to the
relativistic version, the theorem
does neither specify the exact form of the dispersion relation
for small $k$, nor does it determine the number of different
Goldstone modes: these features are not fixed
by symmetry considerations alone -- rather, they depend on the
specific properties of the system.

Turning to the case of Heisenberg ferromagnets,
in a hydrodynamic description
the quadratic form  of
the magnon energies
\begin{equation}\label{ll}
E_k=\frac{\rho_s}{M_0}k^2+{\cal O}(k^4)
\end{equation}
results entirely from the
symmetry of the ferromagnetic state and the fact that
the order parameter is itself a constant
of the motion.\cite{halperin1} Here $M_0$
and $\rho_s$ are, respectively, the magnetization density
and the spin-stiffness constant of the Heisenberg ferromagnet.
This form of the Goldstone modes (which is
just the Landau-Lifshitz formula) can rigorously be obtained by
simple sum-rule arguments (see, e.g., Ref. \onlinecite{keffer}).
Similar arguments are applicable to Heisenberg ferrimagnets
as well.\cite{wolf,hohenberg} An alternative approach,
 relying on the  suggested conformal invariance of
 the model (\ref{ham}), also
predicts the quadratic form of Eq.\ (\ref{ll}).\cite{alcaraz}
Due to the doubling of elementary cell in
the ferrimagnetic ground state,
there appears a second spin-wave branch in the ferrimagnetic case
(optical magnons) with a finite energy gap
$\Delta$ at $k=0$: $\Delta
\sim M_0/\chi_{||}$, where $\chi_{||}$ is the magnetic
susceptibility parallel to ${\bf M}$.\cite{halperin2}
At intermediate temperatures,
it is  the optical magnon branch which produces a number of
specific thermodynamic properties characteristic especially
for the ferrimagnetic systems.
\cite{hagiwara,pati2,yamamoto1,maisinger,kolezhuk1}

The existence of a macroscopic magnetic ground state
also opens an interesting and rare opportunity to apply
the spin-wave theory (SWT) to the low-dimensional
quantum spin system (\ref{ham}). Indeed, recently published
analysis based on the linear spin-wave approximation
\cite{pati2,pati1,brehmer1} qualitatively
confirmed the expected structure of low-energy excitations
in Heisenberg ferrimagnets. Moreover, the second-order
SWT was shown\cite{ivanov2} to produce precise quantitative
results for the ground-state
energy $E_0$ and the on-site magnetizations
$m_1=M_A/N$ and $m_2=M_B/N$
even  for the extreme quantum system $(S_1,S_2)=(1,\frac{1}{2})$:
the second-order SWT results differ by less than $0.017\%$  for $E_0$, and
by less than $0.18\%$ for $m_1$, from the density matrix
renormalization group (DMRG) estimates.\cite{pati2}
In this respect, an interesting question is  to what extend
the spin-wave approach can effectively be used to describe
the properties of the 1D model (\ref{ham}) at zero temperature.
A purpose of the present paper is to  demonstrate,
through an explicit study of the magnon-dispersion perturbation
series, that
the spin-wave approach  can produce precise quantitative results
for the magnon dispersions as well.

The paper is organized as follows. In Section 2, using the
Dayson-Maleev boson representation of spin operators,
the original spin Hamiltonian (\ref{ham}) is transformed to an
equivalent boson Hamiltonian.
The choice of an appropriate  zeroth-order quadratic Hamiltonian
for the perturbation series is also discussed here.
In Section 3 we study the  second-order corrections
for the magnon energies and the related corrections for
the spin-stiffness constant and the optical magnon gap.
Third-order self-energy diagrams and the respective
corrections for the optical magnon gap are also considered
here. In the last Section the results are summarized
and a comparison with available numerical estimates is made.
\section{Dyson-Maleev formalism}
To develop a spin-wave theory
(see, e.g., Refs.\ \onlinecite{nakamura,harris}
and references therein),
one firstly transforms the original spin Hamiltonian
(\ref{ham}) to an equivalent boson Hamiltonian.
We adopt the following Dyson-Maleev representation of
spin operators:
\begin{eqnarray}
S_1^+(i)&=&\sqrt{2S_1}\left(a_i-\frac{1}{2S_1}a_i^{\dag}a_ia_i\right)\,
,\\
S_1^-(i)&=&\sqrt{2S_1}a_i^{\dag},\hspace{0.1cm} S_1^z(i)=S_1-a_i^
{\dag}a_i\, ,\\
S_2^+(j)&=&\sqrt{2S_2}\left(b_j^{\dag}-\frac{1}{2S_2}b_j^{\dag}b_j^
{\dag}b_j\right)\, ,\\
S_2^-(j)&=&\sqrt{2S_2}b_j,\hspace{0.3cm} S_2^z(j)=b_j^{\dag}b_j-S_2\, ,
\end{eqnarray}
where $S^{\pm}_{\alpha}=S_{\alpha}^x\pm
\imath S_{\alpha}^y$, $\alpha =1,2$. $a_i$ and $b_j$ are
boson operators defined on the lattice sites $i \in \cal A$ and
$j \in \cal B$, respectively.

Substituting the  latter expressions in Eq.\ (\ref{ham}), one finds
the following boson representation of ${\cal H}$ in terms
of the Fourier transforms $a_k$ and $b_k$ of $a_i$ and $b_j$:
\begin{equation}\label{ham1}
{\cal H}_B=-2\sigma S^2NJ+{\cal H}_0^{'}+V_{DM}^{'}\,  ,
\end{equation}
where
\begin{equation}\label{ham0}
{\cal H}_0^{'} = 2SJ\sum_k\left[ a_k^{\dag}a_k+\sigma b_k^{\dag}b_k
+\sqrt{\sigma }\gamma_k \left( a_k^{\dag}b_k^{\dag}+a_kb_k\right) \right]\, ,
\end{equation}
and
\begin{eqnarray}\label{dm1}
&V&_{DM}^{'}=-\frac{J}{N}\sum_{1-4}\delta_{12}^{34}\times  \\
&(& 2 \gamma_{1-4}a_3^{\dag}a_2b_1^{\dag}b_4
 +\sqrt{\sigma}\gamma_{1+2-4}a_3^{\dag}
b_2^{\dag}b_1^{\dag}b_4+\frac{1}{\sqrt{\sigma}}\gamma_4a_3^
{\dag}a_2a_1b_4).\nonumber
\end{eqnarray}
$\gamma_k=\cos (ka_0)$ is the lattice structure factor
and $\delta_{12}^{34}
\equiv \delta (1+2-3-4)$
is the Kronecker $\delta$-function. We have used
the convention $(k_1,k_2,k_3,k_4)\equiv (1,2,3,4)$ and
the notations $\sigma \equiv S_1/S_2>1$ and
$S_2\equiv S$. Here and in what follows the external
magnetic field is $H=0$.

In the above expressions ${\cal H}_0^{`}={\cal O}(S)$
is the quadratic boson Hamiltonian
of the  linear spin-wave theory (LSWT) and
$V_{DM}^{'}={\cal O}(1)$ is the quartic Dyson-Maleev
boson interaction. Omitting completely the boson interaction
$V_{DM}^{'}$, a number  of authors has recently used
the quadratic LSWT Hamiltonian
${\cal H}_0^{`}$ for a qualitative description
of the 1D ferrimagnetic
model (\ref{ham}) at zero
temperature.\cite{pati2,pati1,brehmer1}  ${\cal H}_0^{`}$
is easily diagonalized by use of the  following
Bogoliubov
transformation to  the quasiparticle boson
operators $\alpha_k$ and
$\beta_k$:
\begin{eqnarray} \label{uv}
a_k&=&u_k(\alpha_k-x_k\beta_k^{\dag})\, , \hspace{0.3cm}
b_k=u_k(\beta_k-x_k\alpha_k^{\dag})\, ,\nonumber\\
u_k&=&\sqrt{\frac{1+\varepsilon_k}{2\varepsilon_k}}\, ,\hspace{0.3cm}
x_k= \frac{\eta_k}{1+\varepsilon_k}\, ,
\end{eqnarray}
where $\varepsilon_k=\sqrt{1-\eta_k^2}$
and  $\eta_k=2\sqrt{\sigma}\gamma_k/(1+\sigma)$.

As a matter of fact, for the 1D ferrimagnetic model (\ref{ham})
the quadratic Hamiltonian
(\ref{ham0}) is not the most appropriate choice
of a starting zeroth-order approximation
for  perturbation series.
Since we are also interested  in systems with small site spins,
it is more appropriate from the very beginning
to redefine ${\cal H}_0^{'}$ by adding  the first-order $1/S$
corrections to magnon dispersions.\cite{ivanov3} The latter corrections come
entirely from a normal ordering
of the quasiparticle boson operators $\alpha_k$ and $\beta_k$
in the quartic interaction $V_{DM}^{'}$.
 It is important
that these corrections (similar to Oguchi's corrections in
antiferromagnets\cite{oguchi})
renormalize the magnon excitation
spectra (and the ground-state energy)
without changing  their basic  structure, i.e., the number of
Goldstone modes. Unlike the case of Heisenberg antiferromagnets,
where Oguchi's corrections
are numerically small even for site spins  $S=\frac{1}{2}$,
in  quantum 1D ferrimagnets the magnon dispersions
are significantly renormalized\cite{ivanov3} (see below).
As a result of the normal-ordering procedure,
the boson Hamiltonian (\ref{ham1}) is recast to the
following basic form:
\begin{equation}\label{hamb}
 {\cal H}_B =E_0+{\cal H}_0+\lambda V\, ,
 \hspace{0.5cm}   V=V_2+V_{DM}\, ,
 \hspace{0.5cm}\lambda =1\, .
\end{equation}
The ground-state energy $E_0$ calculated
up to first order in $1/S$ reads
\begin{eqnarray}\label{e0}
\frac{E_0}{2NJ}&=&-\sigma S^2+S\left[2\sqrt{\sigma}a_1 +(\sigma
+1)a_2\right]\nonumber \\
&-&a_1^2-a_2^2-\frac{\sigma +1}{\sqrt{\sigma}}a_1a_2\, ,
\end{eqnarray}
where
\begin{equation}
a_1=-\frac{\sqrt{\sigma}}{2(\sigma +1)}\frac{1}{N}
\sum_k \frac{\gamma_k^2}{\varepsilon_k}\, ,\hspace{0.5cm}
a_2=-\frac{1}{2}+\frac{1}{2N}\sum_k \frac{1}{\varepsilon_k}\, .
\end{equation}
Expressed in terms of quasiparticle operators,
the corrected LSWT Hamiltonian reads
\begin{equation}\label{ham00}
{\cal H}_0=2SJ\sum_k\left[
\omega_k^{(\alpha)}\alpha_k^{\dag}\alpha_k
+\omega_k^{(\beta)}\beta_k^{\dag}\beta_k\right] \, ,
\end{equation}
where
\begin{equation}\label{eab}
\omega_k^{(\alpha,\beta)} =
\left(1-\frac{a_1}{S\sqrt{\sigma}}\right)
\left(\frac{\sigma+1}{2}\varepsilon_k\mp
\frac{\sigma-1}{2}\right)-
a_2\frac{1-\gamma_k^2}{\varepsilon_k S}\, .
\end{equation}

The normal-ordered quasiparticle interaction $V$
contains a quadratic term,
\begin{equation}
V_2=J\sum_k\left[V_k^{(+)}\alpha_k^{\dag}\beta_k^{\dag}
+V_k^{(-)}\alpha_k\beta_k \right]\, ,
\end{equation}
where
\begin{equation}
V_k^{(+,-)}=-\frac{a_2(\sigma -1)^2}{\sqrt{\sigma}(\sigma +1)}\left(
1\pm \frac{\sigma +1}{\sigma
-1}\varepsilon_k\right)\frac{\gamma_k}{\varepsilon_k}\, ,
\end{equation}
and the quartic normal-ordered Dyson-Maleev interaction
$V_{DM}$ containing nine vertex functions: $V^{(i)}=V^{(i)}_{12;34}$,
$i=1,\ldots ,9$. 
Explicit expressions for
$V_{DM}$ and the related vertex functions are
presented in   Appendix A.

The  quadratic Hamiltonian (\ref{ham00})
describes two branches of spin-wave excitations (magnons)
defined with  the dispersion relations
$E_k^{(\alpha,\beta)}= 2SJ\omega_k^{(\alpha,\beta)}$, Eq.\
(\ref{eab}). The $\alpha$ excitations are gapless
acoustical  magnons in the subspace
with  $S^z_{tot}=(S_1-S_2)N-1$, whereas
the $\beta$ excitations are gapful optical magnons
in the subspace $S^z_{tot}=(S_1-S_2)N+1$: ${\bf S}_{tot}=\sum_{n=1}^N
\left[{\bf S}_1(n)+{\bf S}_2(n)\right]$.
It is easy to show that the effect of the external magnetic field
in Eq.\ (\ref{ham}) on the dispersions
$E_k^{(\alpha,\beta)}$ reduces to the
simple substitution $E_k^{(\alpha,\beta)} \longmapsto
E_k^{(\alpha,\beta)} \pm g\mu_BH$ provided that $g_1=g_2\equiv g$.

 The spin stiffness constant $\rho_s$
(related to the acoustical branch) plays a basic role (together
with the magnetization density $M_0$) in the
low-temperature thermodynamics of
the model (\ref{ham})\cite{ivanov1,yamamoto1,read,takahashi}
and can be obtained  from the Landau-Lifshitz
relation (\ref{ll}) and Eq.\ (\ref{eab}):
\begin{equation}\label{rho0}
\rho_s^{(0)}=Ja_0S_1S_2\left[ 1-\frac{a_1}{S\sqrt{\sigma}}-
a_2\frac{\sigma +1}{S\sigma}\right]\, .
\end{equation}
In the same free-quasiparticle approximation,
the spectral gap of optical magnons  at $k=0$
reads\cite{ivanov3}
\begin{equation}\label{delta0}
\Delta_0=2J(S_1-S_2)\left(1-\frac{a_1}{S\sqrt{\sigma}}\right)\, .
\end{equation}
In the case $(S_1,S_2)=(1,\frac{1}{2})$
Eqs. (\ref{rho0}) and (\ref{delta0})
give $\rho_s^{(0)}/Ja_0S_1S_2=0.761$ and $\Delta_0=1.676J$.
On the other hand, the LSWT Hamiltonian ${\cal H}_0^{'}$
produces the parameters of the related classical system:
$\rho_s^{(0)}/Ja_0S_1S_2=1$ and $\Delta_0=J$.
The  numerical estimate for the gap
$\Delta=1.759J$\cite{yamamoto2} clearly demonstrates
the importance of the $1/S$ corrections in Eq.\ (\ref{eab}).

Next, let us consider the macroscopic
sublattice magnetizations\cite{ivanov1}
$m_1=\langle S^z_1(i)\rangle =
S_1-\langle a_i^{\dag}a_i\rangle$ and
$m_2=\langle S^z_2(j)\rangle =
-S_2+\langle b_j^{\dag}b_j\rangle$ which are finite
in the ferrimagnetic state. In the case of small site spins and for small
magnetization densities $M_0$,
$m_1$ and $m_2$ are strongly reduced
as compared to their  classical values
(respectively, $S_1$ and $-S_2$):
\begin{equation}\label{m1}
m_1=S_1-a_2\, , \hspace{0.5cm} m_1+m_2=S_1-S_2\, .
\end{equation}
As an example, for $(S_1,S_2)=(1,\frac{1}{2})$
Eq.\ (\ref{m1}) predicts a $42\%$ reduction of the small spin
$S_2=\frac{1}{2}$.
However, the off-diagonal quadratic interaction $V_2$
in Eq.\ (\ref{hamb}) produces important
first-order corrections for the sublattice magnetizations
$m_1$ and $m_2$.
Thus, to first order in $1/S$  Eq.\ (\ref{m1})
should be replaced by the more precise expression
\begin{equation}\label{m11}
m_1=S_1-a_2-\frac{\sqrt{\sigma}}{2S(\sigma +1)^2}\frac{1}{N}
\sum_k\left(
V_k^++V_k^-\right) \frac{\gamma_k}{\varepsilon_k^2}\, .
\end{equation}
Now for the ferrimagnetic chain $(1,\frac{1}{2})$ the
last expression gives the result
$m_1=0.816$  which differs by only  $3\%$ from the DMRG numerical
estimate $0.79248$\cite{pati2}.

Summarizing, it may be stated that the free-quasiparticle
approximation based on the
Hamiltonian (\ref{ham00}) gives  a good qualitative
description of the ground-state properties of
the  model (\ref{ham}).
Further improvement of the SWT results may be
achieved by considering
the role of quasiparticle interactions.
\section{Role of the quasiparticle interactions}
Since the first-order $1/S$ corrections
for the ground state energy $E_0$ and the dispersions
$E_k^{(\alpha,\beta)}$
have already been taken into account by the normal-ordering
procedure, corrections
to Eqs. (\ref{e0}) and (\ref{eab})  arise
only up from the second order of the perturbation series in $V$.
\subsection{Second-order corrections for the magnon energies}
\begin{figure}
\centering\epsfig{file=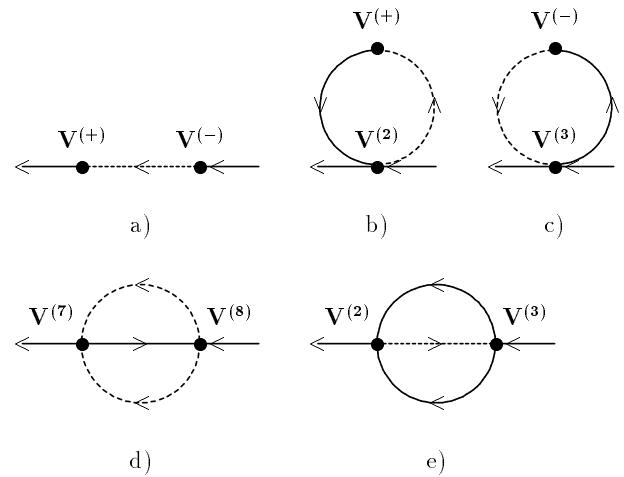,width=8.6cm}
\vspace{0.3cm}
\narrowtext
\caption{Second-order self-energy diagrams giving
the corrections for the dispersion of acoustical magnons
$\omega_k^{(\alpha)}$. Solid and dashed lines represent,
respectively, the propagators of $\alpha$ and $\beta$
magnons. The Dyson-Maleev vertex functions $V^{(i)}$,
$i=1,\dots, 9$ are defined in Appendix A.
}
\label{diag2a}
\end{figure}

The second-order corrections
for the dispersion of
acoustical magnons
$\omega^{(\alpha)}_k$, Eq.\ (\ref{eab}),
are connected with the self-energy
diagrams in Fig.~\ref{diag2a}. The respective
analytic expressions read
\begin{equation}\label{a1}
\delta \omega^{(\alpha)}_k(a)
= -\frac{1}{(2S)^2}\frac{V^{(+)}_{k}V^{(-)}_{k}}
{\omega^{(\alpha )}_k+\omega^{(\beta )}_k }\, ,
\end{equation}
\begin{equation}\label{a23}
\delta \omega^{(\alpha)}_k(bc)=\frac{1}{(2S)^2}\frac{2}{N}
\sum_{p}\frac{V^{(+)}_{p}V^{(2)}_{kp;pk}+V^{(-)}_{p}V^{(3)}_{kp;pk}}
{\omega^{(\alpha )}_p+\omega^{(\beta )}_p }\, ,
\end{equation}
\begin{equation}\label{a4}
\delta \omega^{(\alpha)}_k(d)=-\frac{1}{(2S)^2}\frac{2}{N^2}
\sum_{2-4}\delta_{k2}^{34} \frac{V^{(8)}_{43;2k}V^{(7)}_{k2;34}}
{\omega^{(\alpha )}_k+\omega^{(\alpha )}_2+\omega^{(\beta )}_3+
\omega^{(\beta )}_4 }\, ,
\end{equation}
\begin{equation}\label{a5}
\delta \omega^{(\alpha)}_k(e)=-\frac{1}{(2S)^2}\frac{2}{N^2}
\sum_{2-4}\delta_{k2}^{34}
\frac{V^{(3)}_{43;2k}V^{(2)}_{k2;34}}
{-\omega^{(\alpha )}_k+\omega^{(\beta )}_2+\omega^{(\alpha )}_3+
\omega^{(\alpha )}_4 }\, .
\end{equation}
Note that since the vertex functions $V_k^{(-)}$, $V^{(2)}_{kp;pk}$,
$V^{(3)}_{kp;pk}$, $V^{(8)}_{43;2k}$, and $V^{(3)}_{43;2k}$
vanish at the zone center $k=0$ (see  Appendix A),
the gapless structure of $\omega_k^{(\alpha)}$ is
preserved by each of the second-order corrections,
Eqs.\ (\ref{a1}) to (\ref{a5}).

\begin{figure}
\centering\epsfig{file=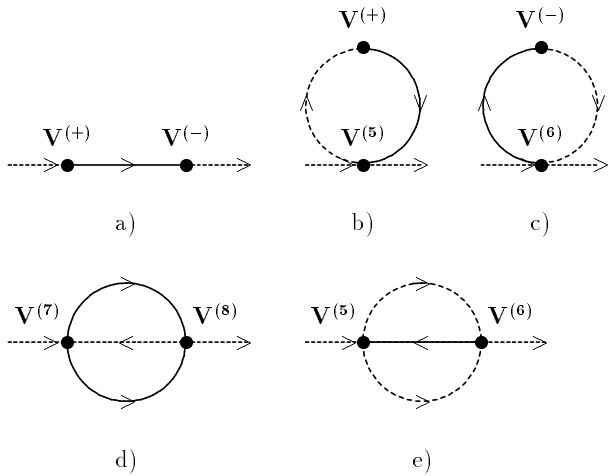,width=8.6cm}
\vspace{0.3cm}
\narrowtext
\caption{Second-order self-energy diagrams giving
the corrections for the dispersion of optical magnons
$\omega_k^{(\beta)}$ (see also the notations in Fig.~\ref{diag2a}).
}
\label{diag2b}
\end{figure}

The second-order corrections
for the dispersion
of optical magnons
$\omega^{(\beta)}_k$, Eq.\ (\ref{eab}), come from similar
diagrams (see Fig.~\ref{diag2b}). The first diagram gives
$\delta \omega^{(\beta)}_k(a)=\delta \omega^{(\alpha)}_k(a)$,
whereas the other
four diagrams give the following contributions:
\begin{equation}\label{bc}
\delta \omega^{(\beta)}_k(bc)=\frac{1}{(2S)^2}\frac{2}{N}
\sum_{p}\frac{V^{(+)}_{p}V^{(5)}_{kp;pk}+V^{(-)}_{p}V^{(6)}_{kp;pk}}
{\omega^{(\alpha )}_p+\omega^{(\beta )}_p }\, ,
\end{equation}
\begin{equation}\label{d}
\delta \omega^{(\beta)}_k(d)=-\frac{1}{(2S)^2}\frac{2}{N^2}
\sum_{2-4}\delta_{k2}^{34}\frac{V^{(7)}_{43;2k}V^{(8)}_{k2;34}}
{\omega^{(\beta )}_k+\omega^{(\beta )}_2+\omega^{(\alpha )}_3+
\omega^{(\alpha )}_4 }\, ,
\end{equation}
\begin{equation}\label{e}
\delta \omega^{(\beta)}_k(e)=-\frac{1}{(2S)^2}\frac{2}{N^2}
\sum_{2-4}\delta_{k2}^{34}
\frac{V^{(5)}_{43;2k}V^{(6)}_{k2;34}}
{-\omega^{(\beta )}_k+\omega^{(\alpha )}_2+\omega^{(\beta )}_3+
\omega^{(\beta )}_4 }\, .
\end{equation}

Note that in the present perturbation
scheme (using ${\cal H}_0$ as a zeroth-order Hamiltonian),
in Eqs.\ (\ref{a1}) to (\ref{e})
there appear the renormalized dispersions $\omega^{(\alpha,\beta)}_k$.
The standard ${\cal O} (1/S^2)$ corrections to
the magnon dispersions can easily be obtained by substituting
in Eqs.\ (\ref{a1}) to (\ref{e}) the bare  excitation energies
(i.e., the functions $\omega^{(\alpha,\beta)}_k$ without
the $1/S$ corrections). Since we are also interested in
the extreme quantum systems which are composed of small site spins
and have small magnetization densities $M_0$
[such as $(1,\frac{1}{2})$ and $(\frac{3}{2},1)$], it may be
expected that the adopted perturbation scheme, where the quasiparticle
interaction $V$ is treated as a small perturbation, is
more appropriate. Such a viewpoint is similar to
Oguchi's treatment of the Heisenberg antiferromagnet\cite{oguchi} and
is supported by the following observations: (i) the higher-order corrections
to the principle approximation for $\omega^{(\alpha,\beta)}_k$
are numerically small (see below), and (ii) the third-order series
in $V$ gives a somewhat better result for the gap $\Delta$
in the extreme quantum system $(1,\frac{1}{2})$.
As a matter of fact,  noticeable deviations from the standard $1/S$
expansions appear only in the above mentioned extreme quantum cases.

\begin{figure}
\centering\epsfig{file=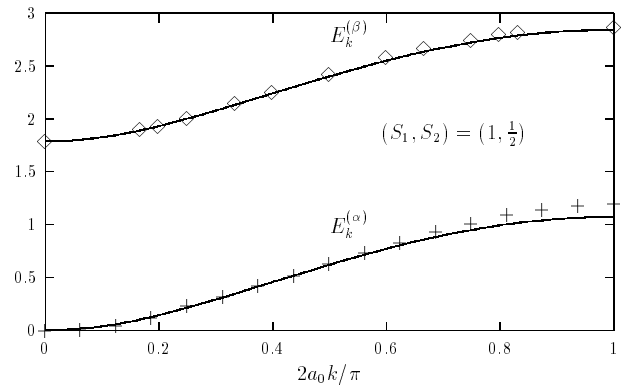,width=8.6cm}
\vspace{0.3cm}
\narrowtext
\caption{Dispersions of the acoustical $E_k^{(\alpha)}$
and optical $E_k^{(\beta)}$ magnons (in $J$ units) in the system
$(S_1,S_2)=(1,\frac{1}{2})$ calculated up to second order in
the magnon interaction $V$. The points ($+$ and $\diamond$) denote,
respectively, quantum Monte Carlo
and exact-diagonalization results (Ref.\ {\protect
\onlinecite{yamamoto2}}).
}
\label{disp}
\end{figure}

The dispersion functions $E_k^{(\alpha,\beta)}$
for the system $(1,\frac{1}{2})$,
calcutated up to second order
in $V$, are presented in Fig.~\ref{disp}. For the optical magnon branch
we find an excellent
agreement with the exact-diagonalization results\cite{yamamoto2}
in the whole Brillouin zone.  Recently, a very successful description of
the optical magnon branch of the system $(1,\frac{1}{2})$
has also been achieved through the
variational matrix product approach.\cite{kolezhuk1}.
As to the acoustical  branch, the agreement with the available
quantum Monte Carlo results\cite{yamamoto2} is not
so satisfactory, especially near the
zone boundary at $k=\pi/2a_0$. It is worth noticing that
in the above region the calculated second-order
corrections to $\omega^{(\alpha)}_k$,
Eq.\ (\ref{eab}), are very small (about $0.5\%$ of the
principal approximation). In principle, it is not
excluded that the spin-wave expansion partially fails to
describe the acoustical branch in the system $(1,\frac{1}{2})$,
as the latter is characterized by the minimal magnetization
density $M_0=1/4a_0$ (see below). To clearify the problem,
further studies are needed in this direction.

The above second-order corrections to
$\omega^{(\alpha,\beta)}_k$ can  be used to
find the coefficient $r_2$ in the perturbation series
for the spin-stiffness constant $\rho_s$
\begin{equation}\label{rho}
\frac{\rho_s}{Ja_0S_1S_2}=1+\frac{r_1}{2S}+\frac{r_2}{(2S)^2}
+{\cal O}\left( \frac{1}{S^3} \right)
\end{equation}
 and the coefficient $\delta_2$ in
the series for the optical magnon gap
\begin{equation}\label{delta}
\frac{\Delta}{2J(S_1-S_2)}=1+\frac{\delta_1}{2S}+\frac{\delta_2}{(2S)^2}
+\frac{\delta_3}{(2S)^3}+{\cal O}\left( \frac{1}{S^4} \right)\, .
\end{equation}
Here $r_1=-2a_1/\sqrt{\sigma}-2a_2(\sigma+1)/\sigma$ and
$\delta_1=-2a_1/\sqrt{\sigma}$ are obtained from
Eqs.\ (\ref{rho0}) and (\ref{delta0}), respectively.
The results for $r_2$ and $\delta_2$ are presented in Table~1.

\subsection{Third-order corrections for the optical  magnon gap}
\begin{figure}
\centering\epsfig{file=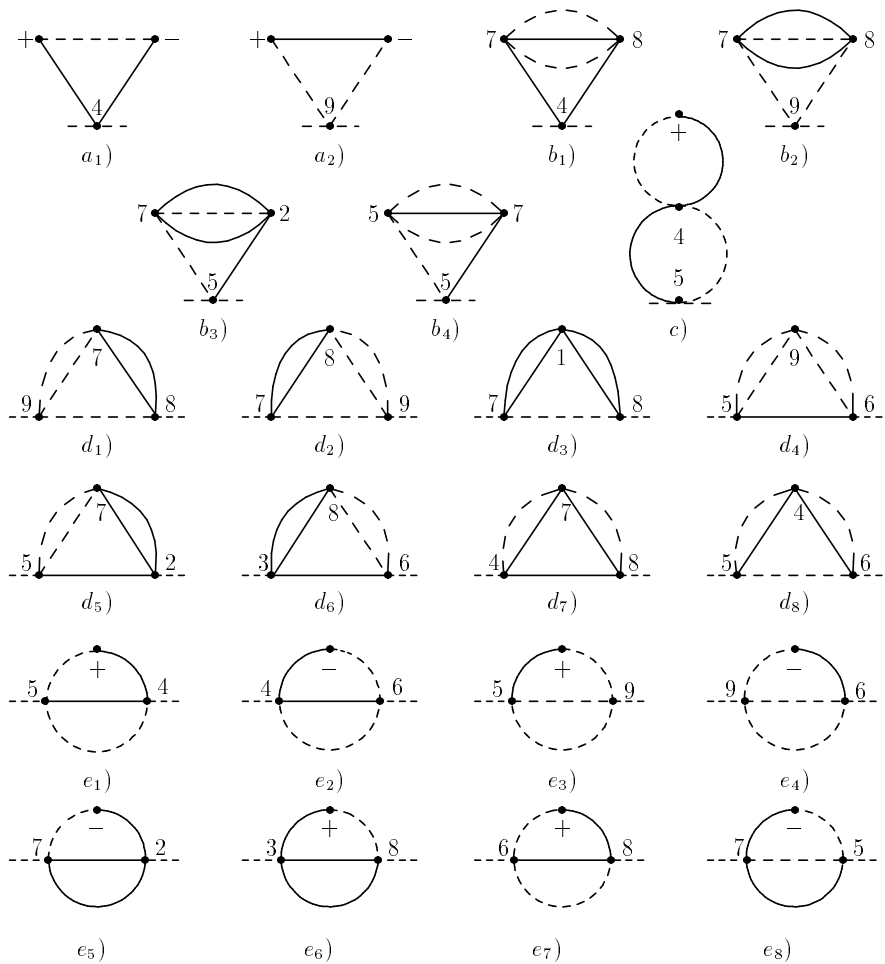,width=12cm}
\vspace{0.3cm}
\caption{Self-energy diagrams giving
the third-order corrections for the optical magnon gap.
The vertices are denoted by their superscripts. The
diagrams $b_1$, $b_3$, $c$, $d_7$, and $d_8$ represent
groups of diagrams which can be obtained by the following
vertex substitutions: $b_1$) (4,8,7) $\mapsto$ (6,6,8);
$b_3$) (5,2,7) $\mapsto$ (6,3,8); $c$) (5,4,+) $\mapsto$ (6,4,-),
(6,8,+) (two diagrams), (5,7,-) (two diagrams);
$d_7$) (4,8,7) $\mapsto$ (7,8,4), (7,4,8); $d_8$) (5,6,4) $\mapsto$
(6,6,8), (5,5,7)(see also the notations in Fig.~\ref{diag2a}).
}
\label{diag3}
\end{figure}

The third-order corrections
for the optical magnon gap $\Delta$ are connected with
the self-energy diagrams in Fig.~\ref{diag3}. These can
be obtained by drawing
all connected  irreducible self-energy diagrams
containing three vertex functions and   only oppositely-oriented
($\alpha$, $\beta$) pairs of internal
magnon lines. The diagrams
containing magnon lines closing
on themselves vanish
since the vertices have already
been normal ordered. In addition, the diagrams
which have internal $\alpha$ lines
carrying the  momentum $k$ give zero contributions as well,
since all vertex functions containing in-going $\alpha$ lines
with momentum $k$ vanish at $k=0$ (see Appendix A).
\end{multicols}
\widetext

The analytic expressions related to the diagrams in Fig.~\ref{diag3}
can easily be obtained by means of standard
diagrammatic rules (see, e.g., Ref.\ \onlinecite{baym}).
For example, the diagrams $a_1$, $d_1$, and
$e_1$ in Fig.~\ref{diag3} give the following contributions for
the dispersion of optical magnons:
\begin{equation}\label{da1}
\delta \omega_k^{(\beta)}(a_1)
= -\frac{1}{(2S)^3}\frac{2}{N}\sum_{p}\frac{V^{(+)}_{p}
V^{(4)}_{pk;pk}V^{(-)}_{p}}
{\left[\omega^{(\alpha )}_p+\omega^{(\beta )}_p\right]^2 }\, ,
\end{equation}
\begin{equation}\label{dd1}
\delta \omega_k^{(\beta)}(d_1)
= -\frac{1}{(2S)^3}\frac{2}{N^3}\sum_{1-5}\delta_{12}^{34}\delta_{12}^{k5}
\frac{V^{(7)}_{12;34}V^{(9)}_{34;5k}V^{(8)}_{k5;12}}
{\left[\omega^{(\alpha )}_1+\omega^{(\alpha )}_2
+\omega^{(\beta )}_3+\omega^{(\beta)}_4\right]
\left[\omega^{(\alpha )}_1+\omega^{(\alpha )}_2
+\omega^{(\beta )}_k+\omega^{(\beta)}_5\right]
}\, ,
\end{equation}
\begin{equation}\label{de1}
\delta \omega_k^{(\beta)}(e_1)
= \frac{1}{(2S)^3}\frac{4}{N^2}\sum_{2-4}\delta_{k2}^{34}
\frac{V^{(+)}_{4}V^{(5)}_{34;2k}V^{(4)}_{2k;43}}
{\left[\omega^{(\alpha )}_4+\omega^{(\beta)}_4\right]
\left[-\omega^{(\beta )}_k+\omega^{(\alpha )}_2
+\omega^{(\beta )}_3+\omega^{(\beta)}_5\right]
}\, .
\end{equation}

The results for $\delta_3$
in systems with different site spins $(S_1,S_2)$ are collected
in Table~1. It is seen that the third-order corrections for
$\Delta$ are  numerically small (as compared to the second-order
correction) even for the extreme quantum system
$(S_1,S_2)=(1,\frac{1}{2})$.
\begin{multicols}{2}
\section{Summary of the results and discussion}
The spin-wave  results for the spin-stiffness
constant $\rho_s$ and the magnon gap $\Delta$ for a number of combinations
$(S_1,S_2)$  are summarized in Table~1. The results
for $\Delta$ are compared with available DMRG estimates.\cite{ono,note}
We find an excellent  agreement with the numerical estimates\cite{ono}
for a number of systems, the largest deviation (about 0.88$\%$)
being connected with the  system
$(1,\frac{1}{2})$.
\end{multicols}
\widetext
\begin{table}
\caption{
Spin-wave results for the spin stiffness constant $\rho_s$
and optical magnon gap $\Delta$ of quantum Heisenberg
ferrimagnetic chains
containing two different
spins $S_1> S_2$ in the elementary cell.
$(r_1,r_2)$ and $(\delta_1,\delta_2,\delta_3)$ are, respectively,
the coefficients of the spin-wave series for
$r_s=\rho_s/Ja_0S_1S_2$, Eq.\ (\ref{rho}),
and $\delta=\Delta/2(S_1-S_2)$, Eq.\ (\ref{delta}).
$\delta^{'}$ indicates numerical estimates{\protect \cite{ono}}
for the reduced excitation gap  $\delta$. $r_s^{(0)}=
\rho_s^{(0)}/Ja_0S_1S_2$.
}
\label{table1}
\begin{tabular}{c|ccccc|ccccc}
$(S_1,S_2)$&$\delta_1$&$\delta_2$
&$\delta_3$&$\delta$&$\delta^{'}$&$r_1$&$r_2$&$r_s^{(0)}$&$r_s$\\
\tableline
$(1,\frac{1}{2})$&0.6756&0.1095&-0.0107&1.7744&1.76&
-0.2391&0.0283&0.7609&0.7892\\
$(\frac{3}{2},1)$&1.0428&0.4262&0.0812&1.6381&1.63&
-0.4907&0.0184&0.7546&0.7592\\
$(\frac{3}{2},\frac{1}{2})$&0.4013&0.0251&-0.0047&1.4217&1.42&
-0.0959&0.0160&0.9041&0.9201\\
$(2,1)$&0.6756&0.1279&-0.0103&1.3685&1.37&
-0.2391&0.0391&0.8804&0.8902\\
$(2,\frac{1}{2})$&0.2861&0.0095&-0.0018&1.2938&1.29&
-0.0517&0.0083&0.9483&0.9566\\
\end{tabular}
 \end{table}
\begin{multicols}{2}
Since in the extreme quantum cases the
constructed perturbation expansions basically
rely on the suggested smallness of the quasiparticle interaction
$V$, it is instructive to check the self-consistency of the theory
by writing the series for $\rho_s$ and $\Delta$
in terms of the formal parameter $\lambda =1$, Eq.\ (\ref{hamb}).
Using the results from Table 1, the series for the system
$(S_1,S_2)=(1,\frac{1}{2})$ read
\begin{equation}
\frac{\rho_s}{Ja_0S_1S_2}
=0.7609\lambda^0+0.0283\lambda^2
+{\cal O}\left( \lambda^3\right)\, ,
\end{equation}
\begin{equation}
\frac{\Delta}{2J(S_1-S_2)}
=1.6756\lambda^0+0.1095\lambda^2
-0.0107\lambda^3+{\cal O}\left( \lambda^4\right)\, .
\end{equation}
The above expansions explicitly demonstrate that the
quasiparticle interaction $V$ in Eq.\ (\ref{hamb})
introduces numerically small corrections to the zeroth-order
principle approximation
based on the quadratic
quasiparticle Hamiltonian ${\cal H}_0=
{\cal O}(\lambda^0)$, Eq.\ (\ref{ham00}).
Of course, the smallness of corrections by itself
does not ensure a good quality of the spin-wave expansion.
The main weakness of the spin-wave theory is connected with the
assumption that the long-range order is well established: it
includes only transverse spin fluctuations, whereas the
longitudinal spin fluctuations are completely neglected.
In this respect, a typical example arises when spin-wave expansions
are used for magnetic systems near the order-disorder critical  point
where the transverse and longitudinal fluctuations
should be treated  on equal ground. In spite of the fact that the
corrections are  small, the spin-wave series
give unsatisfactory  quantitative results.\cite{chubukov}
On the other hand, as the distance from
the critical point increases
the spin-wave description becomes more and more reliable.
The indicated discrepancy for the acoustical branch
in the extreme quantum system $(1,\frac{1}{2})$ (although the
second-order corrections to the principle approximation
are numerically small)
probably reflects the discussed weakness of the SWT.
As a matter of fact,
taking the parameter $M_0$ as a measure of the distance from
the disordered phase, such a behavior of
the spin-wave series in the $(S_1,S_2)$ family
of  1D quantum Heisenberg ferrimagnets  can also be indicated
for the magnon gap $\Delta$ (see Table~1) and
the parameters $E_0$ and $m_1$ (see Ref.\ \onlinecite{ivanov2}):
the largest deviations from the DMRG results appear in the cases
with minimal  magnetization density, $M_0=1/4a_0$.
However, even in the extreme quantum system $(1,\frac{1}{2})$
the discrepancies are small and the spin-wave expansion produces
precise quantitative results. Concerning the dispersion of
acoustical branch $\omega_k^{(\alpha)}$
(and the related spin-stiffness constant
$\rho_s$), more numerical results are needed to
make a statement about the accuracy
of spin-wave description in this case. We believe, however, that at least
for the systems with $M_0>1/4a_0$ the reported  results for $\rho_s$
closely approximate the true spin-stiffness constants
at zero temperature.
\acknowledgments
This work was supported by the Deutsche Forschungsgemeinschaft
(Grant No. 436BUL 113/106/0) and the Bulgarian Science Foundation
(Grant No. F817/98).
\end{multicols}
\widetext
\appendix
\section{Dyson-Maleev vertices for the ferrimagnetic model}
Using the symmetric form adopted in Ref.\ \onlinecite{canali},
the normal-ordered Dyson-Maleev quasiparticle interaction $V_{DM}$,
Eq.\ (\ref{hamb}), reads
\begin{eqnarray}
V_{DM} &=&- \frac{J}{2N} \sum_{1-4} \delta_{12}^{34}
\Big[
V^{(1)}_{12;34} \alpha^{\dag}_1 \alpha^{\dag}_2 \alpha_3 \alpha_4 +
2V^{(2)}_{12;34} \alpha^{\dag}_1 \beta_2 \alpha_3 \alpha_4+
2V^{(3)}_{12;34} \alpha^{\dag}_1 \alpha^{\dag}_2 \beta^{\dag}_3 \alpha_4
\nonumber \\
&+& 4V^{(4)}_{12;34} \alpha^{\dag}_1 \alpha_3 \beta^{\dag}_4 \beta_2
+ 2V^{(5)}_{12;34} \beta^{\dag}_4 \alpha_3 \beta_2 \beta_1
+2V^{(6)}_{12;34} \beta^{\dag}_4 \beta^{\dag}_3 \alpha^{\dag}_2
\beta_1\nonumber \\
&+& V^{(7)}_{12;34} \alpha^{\dag}_1 \alpha^{\dag}_2 \beta^{\dag}_3
\beta^{\dag}_4+V^{(8)}_{12;34} \beta_1 \beta_2 \alpha_3 \alpha_4+
V^{(9)}_{12;34} \beta^{\dag}_4 \beta^{\dag}_3 \beta_2 \beta_1\Big]\, .
\end{eqnarray}
The explicit form of the symmetric
vertex functions for the ferrimagnetic Heisenberg model (\ref{ham})
is
\begin{equation}
V^{(i)}_{12;34}=u_1u_2u_3u_4{\bar V}^{(i)}_{12;34}\, ,\hspace{0.5cm}
i=1,2,\dots ,9\, ,
\end{equation}
where
\begin{eqnarray}
{\bar V}^{(1)}_{12;34} =&+&\gamma_{1-3}x_1x_3+\gamma_{1-4}x_1x_4
+\gamma_{2-3}x_2x_3+\gamma_{2-4}x_2x_4\nonumber\\
&-&\sigma^{1/2}(\gamma_{1-3-4}x_1x_3x_4+\gamma_{2-3-4}x_2x_3x_4)
-\sigma^{-1/2}(\gamma_1x_1+\gamma_2x_2)\, ,\nonumber\\
{\bar V}^{(2)}_{12;34} =&-&\gamma_{1-3}x_1x_2x_3-\gamma_{1-4}x_1x_2x_4
-\gamma_{2-3}x_3-\gamma_{2-4}x_4\nonumber\\
&+&\sigma^{1/2}(\gamma_{1-3-4}x_1x_2x_3x_4+\gamma_{2-3-4}x_3x_4)
+\sigma^{-1/2}(\gamma_1x_1x_2+\gamma_2)\, ,  \nonumber\\
{\bar V}^{(3)}_{12;34} =&-&\gamma_{1-3}x_1-\gamma_{1-4}x_1x_3x_4
-\gamma_{2-3}x_2-\gamma_{2-4}x_2x_3x_4\nonumber\\
&+&\sigma^{1/2}(\gamma_{1-3-4}x_1x_4+\gamma_{2-3-4}x_2x_4)
+\sigma^{-1/2}(\gamma_1x_1x_3+\gamma_2x_2x_3)\, ,\nonumber \\
{\bar V}^{(4)}_{12;34} =&+&\gamma_{1-3}x_1x_2x_3x_4+\gamma_{1-4}x_1x_2
+\gamma_{2-3}x_3x_4+\gamma_{2-4}\nonumber\\
&-&\sigma^{1/2}(\gamma_{1-3-4}x_1x_2x_3+\gamma_{2-3-4}x_3)
-\sigma^{-1/2}(\gamma_1x_1x_2x_4+\gamma_2x_4)\, ,\nonumber  \\
{\bar V}^{(5)}_{12;34} =&-&\gamma_{1-3}x_2x_3x_4-\gamma_{1-4}x_2
-\gamma_{2-3}x_1x_3x_4-\gamma_{2-4}x_1\nonumber\\
&+&\sigma^{1/2}(\gamma_{1-3-4}x_2x_3+\gamma_{2-3-4}x_1x_3)
+\sigma^{-1/2}(\gamma_1x_2x_4+\gamma_2x_1x_4)\, ,\nonumber   \\
{\bar V}^{(6)}_{12;34} =&-&\gamma_{1-3}x_4-\gamma_{1-4}x_3
-\gamma_{2-3}x_1x_2x_4-\gamma_{2-4}x_1x_2x_3\nonumber\\
&+&\sigma^{1/2}(\gamma_{1-3-4}+\gamma_{2-3-4}x_1x_2)
+\sigma^{-1/2}(\gamma_1x_3x_4+\gamma_2x_1x_2x_3x_4)\, ,\nonumber \\
{\bar V}^{(7)}_{12;34} =&+&\gamma_{1-3}x_1x_4+\gamma_{1-4}x_1x_3
+\gamma_{2-3}x_2x_4+\gamma_{2-4}x_2x_3,\nonumber\\
&-&\sigma^{1/2}(\gamma_{1-3-4}x_1+\gamma_{2-3-4}x_2)
-\sigma^{-1/2}(\gamma_1x_1x_3x_4+\gamma_2x_2x_3x_4)\, ,\nonumber  \\
{\bar V}^{(8)}_{12;34} =&+&\gamma_{1-3}x_2x_3+\gamma_{1-4}x_2x_4
+\gamma_{2-3}x_1x_3+\gamma_{2-4}x_1x_4\nonumber\\
&-&\sigma^{1/2}(\gamma_{1-3-4}x_2x_3x_4+\gamma_{2-3-4}x_1x_3x_4)
-\sigma^{-1/2}(\gamma_1x_2+\gamma_2x_1)\, ,\nonumber      \\
{\bar V}^{(9)}_{12;34} =&+&\gamma_{1-3}x_2x_4+\gamma_{1-4}x_2x_3
+\gamma_{2-3}x_1x_4+\gamma_{2-4}x_1x_3\nonumber\\
&-&\sigma^{1/2}(\gamma_{1-3-4}x_2+\gamma_{2-3-4}x_1)
-\sigma^{-1/2}(\gamma_1x_2x_3x_4+\gamma_2x_1x_3x_4).\nonumber
\end{eqnarray}
\begin{multicols}{2}

\end{multicols}

\begin{references}
\bibitem{note}  Apart from  the quantum Monte Calro result
$\rho_s/Ja_0S_1S_2=0.74$ for the system $(1,\frac{1}{2})$
(Ref.\ \onlinecite{yamamoto2}),
we are not aware of any published
estimetes concerning the spin-stiffness constant $\rho_s$
in 1D quantum  Heisenberg ferrimagnets. It may be expected
that the third-order correction for $\rho_s$ is negative,
but we do not believe that the spin-wave
result could closely approximate the cited estimate.
\end{references}
\end{document}